# Extremely large magnetoresistance in few-layer graphene/boron-nitride heterostructures

Kalon Gopinadhan[1,2], Young Jun Shin[2], Rashid Jalil[3], Thirumalai Venkatesan[2,4], Andre K. Geim[3], Antonio H. Castro Neto[1,4], and Hyunsoo Yang[1,2]

[1]Centre for Advanced 2D Materials and Graphene Research Centre, National University of Singapore, 6 Science Drive 2, 117546 Singapore

[2]Department of Electrical and Computer Engineering, and NUSNNI-Nanocore, National University of Singapore, 117576 Singapore

[3]Centre for Mesoscience and Nanotechnology, University of Manchester, Manchester M12 9PL, UK

[4]Department of Physics, National University of Singapore, 117542 Singapore

**Understanding magnetoresistance, the change in electrical resistance upon an external magnetic field, at the atomic level is of great interest both fundamentally and technologically. Graphene and other two-dimensional layered materials provide an unprecedented opportunity to explore magnetoresistance at its nascent stage of structural formation. Here, we report an extremely large local magnetoresistance of ~ 2,000% at 400 K and a non-local magnetoresistance of > 90,000% in 9 T at 300 K in few-layer graphene/boron-nitride heterostructures. The local magnetoresistance is understood to arise from large differential transport parameters, such as the carrier mobility, across various layers of few-layer graphene upon a normal magnetic field, whereas the non-local magnetoresistance is due to the magnetic field induced Ettingshausen-Nernst effect. Non-local magnetoresistance suggests the possibility of a graphene based gate tunable thermal switch. In addition, our results demonstrate that graphene heterostructures may be promising for magnetic field sensing applications.**

Magnetoresistance sensors are widely used in day to day applications[1], where the magnetoresistance (MR) value is an important figure of merit. For example, in information storage applications, the data are retrieved from a magnetic hard disk with a MR read sensor that is extremely sensitive to low (stray) magnetic fields[2, 3]. Prompted by the huge demand for MR sensors with a high sensitivity, low energy consumption, low cost and ready availability, researchers are investigating various new materials. Graphene, a stack of single layer carbon atoms arranged in a hexagonal periodic lattice with weak van der Waals interlayer interaction, can be a wonder electronic material which can exhibit large MR values. Fundamentally, atomically thin structures provide the simplest system to understand the origin of MR, thanks to the discovery of semi-metallic graphene and other low dimensional conducting structures[4-6].

In the past, attempts have been made to understand the MR in graphene. A linear MR of 80 – 250% was reported in epitaxial multilayer graphene at 2 K in a normal magnetic field of 12 T.[7] The explanation of the temperature independent linear MR based on the quantum theory[8] fails to account for the observed MR at room temperature as no quantum effects are expected at room temperature. A linear and quadratic MR of 60%, at 300 K and a magnetic field of 14 T, is reported in chemical vapor deposition (CVD) grown few-layer graphene with the current perpendicular to the film plane.[9] The authors attribute this large MR to in-plane velocity of the carriers due to disorientation of the graphene atoms, when it is sandwiched to make stacked few-layer graphene. In addition, this structure suffers from defects, grain boundaries and impurities which limit the mobility of the carriers reducing the MR. By increasing the mobility with graphene peeled from Kish graphite[10], a larger room temperature MR can be achieved. However, it is found that highly ordered pyrolytic graphite (HOPG) samples with sizes less than 100 μm exhibit a very small MR[11] and the smaller MR observed in the past in few layer graphene samples is explained by the size effect. Recently, a sizable

MR is reported at 300 K from single layer graphene[12] due to enhanced scattering from charged impurities. A negative MR is also predicted[13] and observed[14] in graphene nanoribbons, but the MR value is limited (~100%). A finite MR has been reported recently on graphene/BN vertical heterostructures[15, 16].

Here we show various scenarios such as local Hall bar, van der Pauw and non-local geometry, where the multilayer graphene can show large MR values. In a local geometry, small-sized (few microns) few-layer graphene samples can exhibit an extremely large MR of 2,000% at 400 K in graphene/boron-nitride (BN) heterostructures. We provide experimental evidence to show that few-layer graphene can be an electronic material for MR sensors at both low and high magnetic fields, and at practical device operating temperatures of upto 400 K, with a very small temperature coefficient of resistance. In addition, we show that the MR is electric field tunable thus providing additional functionalities to the sensor. Furthermore, we show that Ettingshausen - Nernst effect based non-local MR sensors can exhibit MR > 90,000% at room temperature.

**Results**

**Electric and magnetic field effects on channel resistivity.** Based on the multi-channel model (see Supplementary Notes 1-2) by considering few-layer graphene exhibiting parallel transport with different mobilities and densities of the carriers, we find that the MR can be arbitrarily large for certain combinations of mobility and carrier concentration in different channels. The underlying physics is that the MR is due to the Lorentz force on the charge carriers with a cosine dependence on the angle of the magnetic field. When a magnetic field is applied normal to the multilayer graphene plane, there will be parallel conduction between channels, which can be effectively represented by the multi-channel model. In a low magnetic field (**B**), the current flows though the high mobility ($\mu$) channel (because its conductivity is higher), whereas in high **B** its longitudinal conductivity rapidly drops as $1/[1+(\mu B)^2]$ forcing the current towards into the low mobility channel that now has a higher longitudinal conductivity of the channels.

The theoretically predicted MR behaviors with experimentally accessible carrier densities and mobilities are shown in Fig. 1a. It is clear that the MR is very large even for moderate magnetic fields. Among various combinations of mobility and carrier density, it is inferred that a large mobility (which can be achieved by using graphene on BN) with respect to the second channel or a lower carrier concentration (by tuning the gate voltage to near the charge neutrality point where the carrier density is at minimum) is the key to obtain a high magnetic field sensitivity at low magnetic fields. It is known that the intrinsic carrier density at the charge neutrality point (CNP) is of the order of $\sim 10^{10}$–$10^{11}$ cm$^{-2}$ depending on the quality of graphene[17]. The mobility at the CNP is usually more than 10,000 cm$^2$ V$^{-1}$ s$^{-1}$,[10] which means that the MR can be arbitrarily large at the CNP.

To support the model, graphene on BN has been prepared by a two-step mechanical transfer process as described elsewhere.[18] The optical micrograph of the fabricated device with 4 graphene layers is shown in Fig. 1b. The present study aims to understand the properties of few-layer graphene at the practical device temperature of 400 K which has not been reported previously. The source-drain voltage ($V_{sd}$) versus source-drain current ($I_{sd}$) shows a linear characteristic at all gate voltages implying an ohmic contact between graphene and Cr/Au electrodes (Fig. 1c). The back gate voltage ($V_G$) dependence of the channel resistance shows a maximum at ~ 6 V corresponding to the CNP. The CNP is broad which could be related to both the band structure of few-layer graphene and the thermal smearing of the Fermi surfaces at 400 K. In order to understand the MR effect in this layered system, a magnetic field of 9 T is applied normal to the graphene plane. It is clear from Fig. 1d that the applied magnetic field increases the resistance of charge carriers resulting a positive MR at all gate voltages. In line with the model discussed above, a much larger change is observed near the CNP due to a lower carrier concentration and larger mobility.

**MR and Hall data at 400 K.** Figure 2a shows the MR of 4 layer graphene/BN as a function of the external magnetic field at 400 K in various $V_G$. It is clear that the maximum MR is ~880% at 9 T which is near the CNP, the highest value ever reported in graphene at this temperature. At CNP, the MR is parabolic at low magnetic fields, but it is linear at high magnetic fields with no indication of any saturation. As $V_G$ is tuned away from the CNP, the MR percentage decreases on either side. At $V_G$ = -25 V, the MR falls to 130% at 9 T. To check whether BN really improves the MR, similar magneto-transport measurements have been carried out with graphene on $SiO_2$. It is clear from Fig. 2b that the MR is smaller on $SiO_2$ due to the degradation in mobility as previously reported[17]. In the graphene/$SiO_2$ case, the MR is also gate tunable and the characteristics are very similar to that of graphene/BN implying that the intrinsic property of few-layer graphene is responsible for the MR (see Supplementary Figures 1-3). The CNP in this case is shifted slightly to a higher gate voltage than that of graphene/BN which points to the different reactivity of the surface of graphene attached to BN or $SiO_2$ due to a difference in charge inhomogeneity[19].

To better understand the MR characteristics, the Hall resistivity $\rho_{xy}$ has been measured at different $V_G$, as shown in Fig. 2c. The Hall resistivity shows a non-linear behavior at various $V_G$ suggesting contributions from different layers with different carrier mobilities and carrier concentrations. Above the CNP, the $\rho_{xy}$ versus H curve shows a negative slope suggesting that the majority charge carriers are electrons, and below the CNP the majority carriers are holes as evidenced from the positive slope. Both the MR and $\rho_{xy}$ are fitted with a multi-channel model described in Supplementary Notes 1-2. The extracted parameters, carrier density and mobility for $V_G < V_{CNP}$, are shown in Fig. 2d, which are modulated with $V_G$. Note that, in the multi-channel model, the graphene layer close to BN forms channel 1 and the rest of the layers form channel 2. Due to the thermal excitation of carriers at a measurement temperature of 400 K as well as the Ettingshausen-Nernst effect which is discussed later, the fitting with the model works well away from CNP. Typical mobilities of the charge carriers in the two effective channels are $\mu_1$ = ~10,000 (channel 1) and $\mu_2$ = ~1,500 cm$^2$ V$^{-1}$ s$^{-1}$ (channel 2) at $V_G$ = -25 V, and the large mobility difference in the two channels is the reason to obtain a large MR, which is in accordance with the model. The channel 1 close to BN has the highest mobility and the rest of the layers (channel 2) show graphitic mobility. It has also been shown that the bulk mobility of graphite is ~ 3000 cm$^2$ V$^{-1}$ s$^{-1}$,[20] which is in agreement with our extracted mobility value for channel 2. In addition, microscopic disorder is also present in our samples which is evident from the linear MR characteristics as well as the non-saturating MR,[12] which might also be playing a role in inducing the large mobility difference between channel 1 and channel 2.

The carrier density is modulated by $V_G$, however we found different types of carriers in the two channels at a certain $V_G$. For example, when $V_G < V_{CNP}$, the hole density is larger than the electron density, and for $V_G > V_{CNP}$, the electron density is larger than the hole density. This is a result of the electric field modulation of the carrier density in channel 1 close to BN, however the carrier density of channel 2 remains in the bulk graphite compensated state with weak gate voltage tunability. The screening length in multilayer graphene is much debated, and the various theoretical[21, 22] and experimental studies[23, 24] show that depending upon the doping, the screening length can vary from half a unit cell to an order of magnitude. The observed non-linear Hall effect is an indication that the screening length is lower than the multilayer thickness in our samples. It is inferred that the applied $V_G$ modulates charge carriers in 1–2 graphene layers close to BN, whereas the charge density in rest of the graphene layers are not much affected. The hole density of channel 1 reaches to ~

$1.5\times10^{12}$ cm$^{-2}$ at $V_G$ = -25 V with a graphene/BN (10 nm)/SiO$_2$(300 nm) back gate structure. A charge density of $5-7\times10^{10}$ cm$^{-2}$ per V is expected to induce with a back gate. Since the CNP is ~ 10 V, an applied $V_G$ of -25 V induces a hole density of $1.8-2.5\times10^{12}$ cm$^{-2}$ and the extracted hole density of $1.5\times10^{12}$ cm$^{-2}$ suggests that more than 75% carriers are in the first channel. It has been predicted theoretically[22] that for a carrier density of $10^{12}$ cm$^{-2}$, the screening length is 1–2 layers. The non-linear Hall effect clearly indicates the presence of two different carrier types and its modulation on applying $V_G < V_{CNP}$ which is shown in Fig. 2d. The applied electric field presumably shifts the Dirac point in each layer to match the Fermi level as predicted theoretically[22], which could be the reason as to why two different types of carriers are seen. In channel 1, the hole density decreases on increasing $V_G$ from -25 V towards the CNP, whereas the mobility increases which is in agreement with the reported results. In channel 2, the magnitude of electron density and mobility is lower than that of channel 1, and shows a much weaker dependence on $V_G$, which may be the result of screening.

**Temperature and angle effects on MR.** A much larger MR effect is seen in 6 layers graphene/BN samples and the MR behavior at $V_G$ = 0 V is shown in Fig. 3d. The local measurement geometry is shown in Fig. 3a. The estimation of layer number by Raman spectrum is provided in Supplementary Figure 4. The MR at 400 K is ~ 2,000% at 9 T and the MR hardly changes as the sample is cooled to 300 K suggesting a very small temperature coefficient of resistance in this range of temperature. When the sample is further cooled to 2 K, the MR increases to a very large value of ~ 6,000% at 9 T. At first glance this large increase in the MR at a low temperature may suggest the importance of quantum effects, however the Landau levels are only partially formed (as inferred from weak oscillations in resistance on top of a huge classical MR background) at 2 K which argues against purely quantum related contribution to the large observed MR.

To better understand the role of additional contributions to the MR, it has been measured at different magnetic field orientations with respect to the film plane (Fig. 3e). The magnetic field orientation has been changed from out-of-plane ($\theta = 0°$ and $180°$) to in-plane ($\theta = 90°$ and $270°$) in a complete cycle with a constant magnetic field of 9 T. The MR is maximum when the magnetic field is normal to the sample plane which suggests that the Lorentz force $\mathbf{F} = q(\mathbf{v}\times\mathbf{B})$, where $q$ is the charge of the carriers, $\mathbf{v}$ is the in-plane velocity along the current direction, and $\mathbf{B}$ is the strength of the magnetic field, causes deflection of the charged particles in the graphene plane. At 300 and 400 K, the MR follows cosine dependence with angles confirming the role of the Lorentz force. However, at 2 K, the MR deviates from a simple cosine behavior which indicates the presence of additional contributions to the MR. From the cosine fit, it is calculated that the non-classical related MR is only ~ 700% as compared to a classical MR of ~ 5,300%. This study clearly separates the classical and non-classical MR contributions, and the result suggests that the classical MR dominates in graphene even at 2 K, contrary to other reports[7, 14].

In addition, we have performed the MR measurements by changing the orientation of the magnetic field in the film plane (see Supplementary Figure 2b) and found that there is a negligible MR, as the thickness of the graphene layers is too thin to have any Lorentz force effect on the resistance in the case of in-plane fields, further confirming that a classical Lorentz force arising from the interaction of charged particles with the graphene lattice is the dominant mechanism. It is to be noted that the geometry related contribution is negligibly small in this configuration, as evidenced from the simple cosine variation of the MR at high temperatures (300 and 400 K). In addition, the in-plane MR, where both the magnetic field and current are parallel, is less than 1% suggesting that the contribution of inter-layer interactions to the observed MR is negligibly small.

**MR in van der Pauw geometry.** In order to understand the origin of non-classical MR and improve the magnetic field sensitivity of the graphene sensor, measurements of the MR with current electrodes (the injector) separated from voltage electrodes (the detector) have been carried out in the van der Pauw geometry (see Fig. 3b for the measurement geometry and Supplementary Figure 5 for the device image). Surprisingly, the MR measured in this way in Fig. 3f is extremely large reaching ~35,000% at 50 K and 5,000% at 300 K in comparison to a local MR of 2,000% at 300 K. Note that since graphene is not constricted, the classical ohmic contribution is enhanced by a geometric effect, although the non-classical contributions can be still present in this geometry. The large increase in the MR at low temperatures cannot be explained simply by invoking the suppression of a phonon contribution which increases the mobility and hence MR, since the percentage increase in the MR with respect to that at room temperature is much higher than expected. The deviation from cosine dependence at a low temperature (Fig. 3e) indicates that an additional mechanism is present. Of several possibilities, it is reasonable to assume that the spin Hall contribution is negligible, as the intrinsic spin-orbit coupling in graphene is weak due to a low atomic number. We discuss several other possibilities below based on the experimental results.

**Non-local MR.** The possibility of a heat related contribution as the source of non-classical MR is likely. To verify this, we have selected 4 layer graphene (more results are in Supplementary Figures 6-10) with a narrow channel width ($W$) of 7 μm and the separation ($L$) of the non-local voltage electrodes from current electrodes is 21 − 29 μm which yields a $L/W$ ratio ≥ 3. The result of the non-local resistance ($R_{NL}$) and the MR at 300 K with $L$ = 29 μm is shown in Fig. 3g. It is observed that a large non-local MR value of 90,000% exists even at electrode separations where a classical ohmic contribution is not expected. It is to be noted that the ohmic contribution decays exponentially as $\rho_{ohm} = \rho_0 \exp(-\pi L/W)$.[25, 26] The MR signal attains a maximum at the CNP showing that the signal is intrinsic to graphene. To understand the effect of orientation of the magnetic field on the non-local signal, we have performed an angular dependent MR measurement at 300 K under a constant magnetic field of 9 T, as shown in Fig. 3h. The measurement has been performed at $V_G$ = 0 close to the CNP. The orientation of the magnetic field has been varied from out-of-plane ($\theta$ = 0° and 180°) to in-plane ($\theta$ = 90° and 270°), and the results show $R_{NL} \propto \cos\theta$.

Possible candidates for non-local signal are thermally induced classical effects such as the Joule heating, Nernst, and Ettingshausen effects. Joule heating at the injector would cause an energy flow, $Q = I^2R$, where $I$ is the injector current. This energy flow produces a temperature gradient leading to a non-local Nernst voltage at the detector. The Nernst effect is the magnetic analogue of the Seebeck effect[27, 28] whereby a potential difference appears across a sample subject to a temperature gradient in a perpendicular magnetic field. To test this idea, we have varied the injector current ($I$) and measured the non-local voltage which shows a linear characteristic (see Supplementary Figure 7c), excluding Joule heating as a possibility. On the other hand, the Ettingshausen effect is the magnetic analogue of the Peltier effect where charge flow in a sample subject to a magnetic field creates a thermal gradient (Fig. 3c), leading to a non-local voltage at the detector linear in current via the Nernst effect[29]. The measured non-local voltage with the injector current is linear in our sample, which suggests the Ettingshausen-Nernst effect as the source of the non-local MR. Non-local contacts with different $L$ have been examined and all exhibit a linear relationship between currents and non-local voltages. The cosine dependence of $R_{NL}$ with respect to the magnetic field also suggests the Ettingshausen-Nernst effect as the cause, since both Ettingshausen and Nernst effects generate thermal voltages via the Lorentz force which is cosine dependent on the magnetic field.

**Discussion**

A giant non-local signal has been reported in the presence of a magnetic field in single layer graphene at low temperatures and the origin is attributed to the Zeeman spin Hall effect (ZSHE)[30]. In addition, thermal related Nernst and Ettingshausen effects are reported in single layer graphene at low temperatures which results in a non-local signal[28]. We consider the ZSHE and its inverse at the CNP due to combined Zeeman and orbital effects, however the Zeeman splitting energy, $g\mu_B B \sim 1$ meV (for $g = 2$ and the intensity of the applied magnetic field $B = 9$ T) is too small in comparison to the thermal energy (26 meV) to explain a large $R_{NL}$ of ~ 1 k$\Omega$ at 300 K. The ZSHE induced $R_{NL}$ is related to $B_\perp$ (due to orbital effect) and $B_{tot}$ (due to Zeeman effect), where the total applied field $B_{tot}^2 = B_\perp^2 + B_\parallel^2$. Here the orbital effect is a complicated function of geometry and $B_\perp$, whereas $B_{tot}$ is independent of the orientation of the magnetic field. However, the measured non-local signal in Fig. 3h is a simple cosine function of the angle which argues against the ZSHE. It should be noted that the non-local signal is maximum when the magnetic field is normal to the graphene plane which argues against the Zeeman effect where it is independent of the orientation of the magnetic field.

It is clear from the above discussion that the deviation of MR from cosine dependency at 2 K shown in Fig. 3e (as well in Supplementary Figure 2a) is most probably originated from quantum effects. However, its magnitude is much smaller than ohmic and thermomagnetic contributions. In addition, we geometrically separate the contributions from ohmic and thermomagnetic effects, both of which are cosine dependent and unambiguously confirm that classical MR (ohmic and thermomagnetic) contributions are dominant at all temperatures. For the first time, we show that the thermomagnetic effects can be observed even at an elevated temperature of 350 K in few-layer graphene sample (see Supplementary Figure 10a). It is to be noted here that the Ettingshausen and Nernst effects are typically observed in poor thermal conductors like Bismuth[31]. Graphene is reported to have the highest thermal conductivity[32, 33] and thermal gradients were thought to be highly unlikely. However, we show that near the CNP, where the carrier density is at minimum, thermal gradients exist and the thermal conductivity of graphene can be tuned from high to low with a back gate voltage. This opens a possibility of graphene-based thermal switch devices with a gate tunable thermal conductivity.

The stacking has been shown to affect the electrical transport properties of trilayer graphene[34], especially at the CNP and at low temperatures. In a 4 layer graphene, 4 stacking sequences are possible, ABAB (Bernal), ABCA (Rhombohedral), ABAC, and ABCB. Among these, the last two are electronically predicted to be unstable[35]. A theoretical study[36] on 4 layer graphene predicted a semi-metallic band structure for ABAB stacking, whereas for ABCA stacking a gaped band structure is predicted. Since we observe a CNP in our few-layer graphene samples, this is suggestive of a Bernal stacking. We have not made an attempt to differentiate the effect of stacking sequences on MR in this study, and we believe that the stacking sequence can probably change the magnitude of MR via change in interlayer coupling, however the physics will remain the same. Modeling MR with the inclusion of stacking sequence when the layer number is more than three may be computationally challenging as the number of parameters are very large. Nevertheless, we believe that this is a very interesting subject for future studies.

**Methods**

**Device fabrication**. Graphene has been prepared by mechanically peeling Kish graphite and transferred to either heavily doped silicon with a layer (300 nm thick) of $SiO_2$ on top of Si or to BN previously transferred by a dry transfer technique[18]. The samples have been annealed in forming gas at 250 °C in every step of fabrication to improve the interface quality. Electrodes were patterned by e-beam lithography using a bilayer resist combination of MMA/PMMA and subsequently Cr/Au (7 nm/95 nm) was deposited by an e-beam evaporator, where a large distance between the source materials and the sample keeps the damage induced in graphene at minimal. Standard lift-off procedures using warm acetone were followed after the deposition. Optical microscope based contrast is used to get initial information about the number of layers. In order to determine the quality and thickness of graphene, Raman spectroscopy measurements were carried out (see Supplementary Figure 4). A laser wavelength of 532 nm with a power density ~ 0.5 mW $cm^{-2}$ was used to avoid any laser induced heating.

**Transport measurements**. The electrical transport measurements were carried out in a physical property measurement system (PPMS) under helium atmosphere with a source meter (Model 2400, Keithley Inc.) and a multimeter (Model 2002, Keithley Inc.). Before the measurement, the samples have been annealed for 2 hours at 400 K under helium atmosphere to remove any adsorbed water vapor. To apply back gate bias, the source terminal was connected to the back gate and the leakage current through the $SiO_2$ (or BN/$SiO_2$) layer was monitored.

**Acknowledgments**

This research is supported by the National Research Foundation, Prime Minister's Office, Singapore under its Competitive Research Programme (CRP Award No. NRF-CRP6-2010-5) and its Medium Size Centre programme.

**Author contributions**

K.G., A.K.G., A.H.C.N., and H.Y. planned the study. K.G., Y.S., and R.J. fabricated devices. K.G. measured transport properties. A.H.C.N. provided the theoretical work. K.G., T.V., A.K.G., A.H.C.N., and H.Y. discussed the results. K.G. and H.Y. wrote the manuscript. H.Y. supervised the project.

**Additional information**

**Competing financial interests:** The authors declare no competing financial interests.

**FIGURES**

**Figure 1| Electric and magnetic field effect of channel resistivity in few-layer graphene. a**, Resistivity versus external magnetic field ($H$) based on the theoretical prediction for a certain combination of carrier densities and mobilities in two effective channels suggesting the possibility of very high magnetic field sensitivity at moderate magnetic fields. Case I: $n_1=10^{11}$ cm$^{-2}$, $n_2=1.1\times10^{11}$ cm$^{-2}$, $\mu_1=20,000$ cm$^2$ V$^{-1}$ s$^{-1}$, $\mu_2=1,000$ cm$^2$ V$^{-1}$ s$^{-1}$, Case II: $n_1=8\times10^{12}$ cm$^{-2}$, $n_2=1.1\times10^{11}$ cm$^{-2}$, $\mu_1=5,000$ cm$^2$ V$^{-1}$ s$^{-1}$, $\mu_2=1,000$ cm$^2$ V$^{-1}$ s$^{-1}$, and Case III: $n_1=10^{11}$ cm$^{-2}$, $n_2=1.1\times10^{11}$ cm$^{-2}$, $\mu_1=200,000$ cm$^2$ V$^{-1}$ s$^{-1}$, $\mu_2=10,000$ cm$^2$ V$^{-1}$ s$^{-1}$. **b**, Optical micrograph of the fabricated 4 layer graphene on top of BN and SiO$_2$. The location of graphene is indicated by red dashed lines. **c**, The voltage ($V_{sd}$) as a function of source-drain current ($I_{sd}$) at various gate voltages ($V_G$) suggesting an ohmic contact for Cr/Au electrodes on graphene (Gr). **d,** The channel resistivity ($\rho$) as a function of $V_G$ at 400 K. A magnetic field of 9 T is also applied normal to the graphene plane.

**Figure 2| Magnetoresistance and Hall resistivity at 400 K. a**, The magnetoresistance (MR) versus external magnetic field ($H$) of 4 layer graphene/BN as a function of $V_G$ at 400 K. The MR is maximum at the charge neutrality point (CNP ~ 10 V). **b**, The MR versus $H$ of graphene/SiO$_2$ as a function of $V_G$ at 400 K. The charge neutrality point (CNP) is ~ 25 V. **c**, The Hall resistivity ($\rho_{xy}$) as a function of $H$ at various $V_G$, which shows a non-linear behavior due to contribution from various layers. The slope changes upon crossing the CNP suggesting a change in the majority carrier type. The MR contribution arising from any geometry effect is removed from the measured Hall resistivity. **d**, The fitted parameters, carrier density ($n$) and mobility ($\mu$), extracted from the multi-channel model. For $V_G < 0$, positive $n_1$ values indicate hole transports, whereas negative $n_2$ suggests electron transports.

**Figure 3| Angle and temperature dependent magnetoresistance in various geometries. a**, The schematic of the local measurement geometry. **b**, The schematic of the van der Pauw measurement geometry. **c**, Schematic of the non-local geometry which identifies Ettingshausen-Nernst effect. An applied current (*I*) in the *y*-direction and a normal magnetic field in the *z*-direction ($B_z$) generate a charge current in the *x*-direction due to the Ettingshausen effect, which accumulates at the non-local electrode and raises the temperature. Due to the Nernst effect, a voltage is generated at the non-local electrodes ($V_{NL}$) along the *y*-direction. **d**, The magnetoresistance (MR) versus magnetic field (*H*) at various temperatures for a graphene sample with 6 layers on BN. The MR is very large ~ 2,000% even at a practical operating temperature of 400 K. The MR increases to a larger value of 6,000% at 2 K. **e**, The angle dependence of the MR can be fitted with cos($\theta$) at 300 and 400 K (shown as solid black curves) implying a dominant classical MR effect. However, the MR at 2 K cannot be fitted with a simple cosine relationship indicating contributions from other effects. **f**, The van der Pauw geometry MR at various temperatures for graphene on BN. The channel width (*W*) and distance of separation (*L*) of voltage electrodes from the current electrodes is 20 and 7 μm, respectively, yielding a *L/W* ratio of < 1, justifying the definition of van der Pauw geometry. The MR is ~35,000% at 50 K with a higher magnetic field sensitivity. **g**, The non-local MR from a narrow channel 4 layer graphene/BN device at 300 K showing the importance of magnetic field induced Ettingshausen-Nernst effect. The channel width (*W*) and distance of separation (*L*) of voltage electrodes from the current electrodes is 7 and 29 μm, respectively, yielding a *L/W* ratio > 4, justifying the definition of non-local geometry. **h,** The non-local angular MR at 9 T and 300 K suggesting the presence of Ettingshausen-Nernst effect which is cosine dependent.

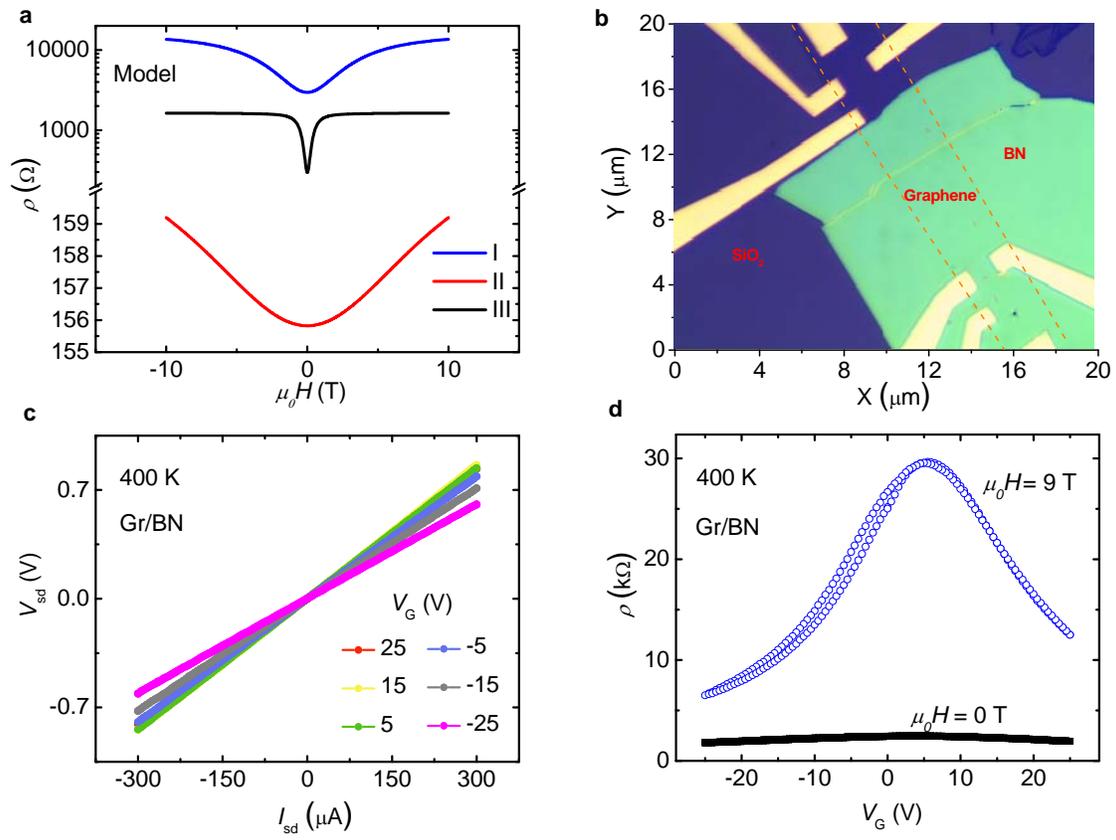

Figure 1

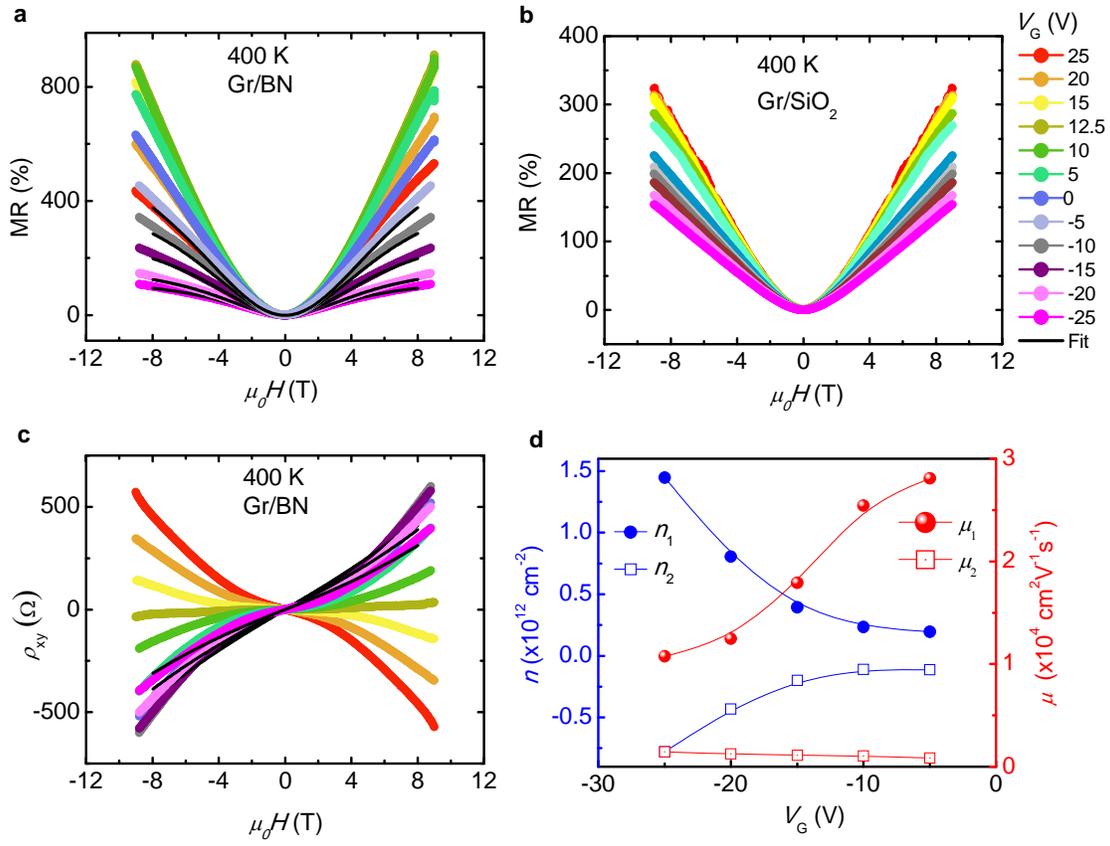

Figure 2

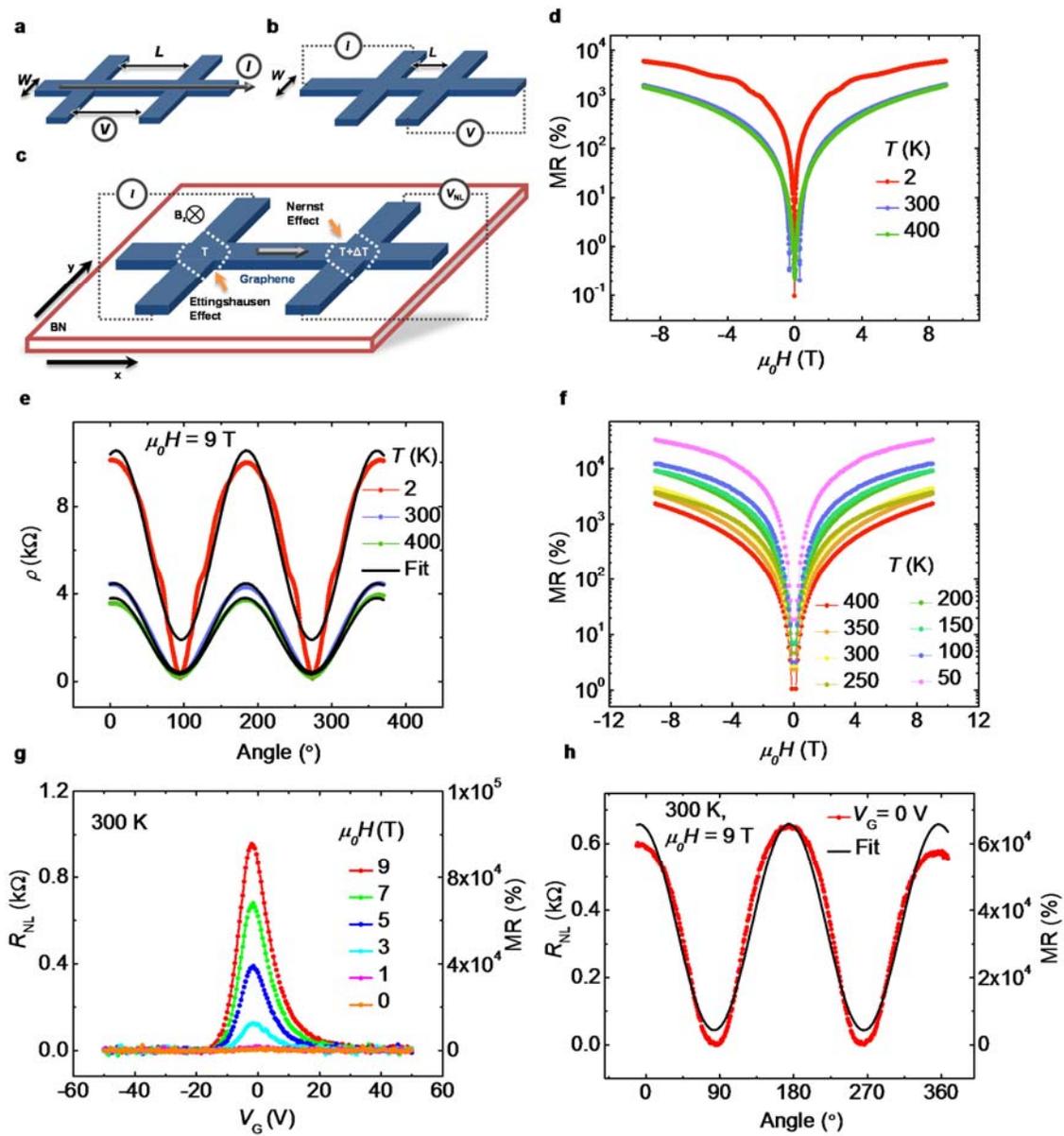

Figure 3